\documentstyle[12pt]{article}
\begin{document}
\title{A Short Range Force}
\author{B.G. Sidharth\\
Centre for Applicable Mathematics \& Computer Sciences\\
Adarsh Nagar, Hyderabad - 500 063, India}
\date{}
\maketitle
\begin{abstract}
Gravitomagnetic and gravitoelectric forces have been studied for
sometime and tests for detecting such forces arising from the
earth, are under way. We apply similar considerations at the level
of elementary particles in a formulation using General Relativity,
and deduce the presence of short range forces. A possible
candidate could be the somewhat recently detected but otherwise
mysterious short range $B_{(3)}$ force, mediated by massive
"photons".
\end{abstract}
\section{Introduction}
In some ways the General Relativistic gravitational field
resembles the electromagnetic field, particularly in certain
approximations, as for example when the field is stationary or
nearly so and the velocities are small. In this case the equations
of General Relativity can be put into a form resembling those of
Maxwell's Theory, and then the fields have been called
Gravitoelectric and Gravitomagnetic \cite{r1,r2}. Experiments have
also been suggested for measuring the Gravitomagnetic
force components for the earth \cite{r3}.\\
We can ask whether such a consideration can be applied to
elementary particles, if in fact they can be considered in the
context of General Relativity. It may be mentioned that apart from
Quantum Gravity, there have been three different approaches for
studying elementary particles via General Relativity
\cite{r4,r5,r6} and references therein. We will now show that it
is possible to extend the Gravitomagnetic and Gravitoelectric
formulations to elementary particles within the framework of the
theory developed in \cite{r6}.
\section{The Elementary Particle Case}
In \cite{r6}, the linearized General Relativistic equations are
seen to describe the properties of elementary particles, such as
spin, mass, charge and even the very Quantum Mechanical anomalous
gyromagnetic ratio $g = 2$, apart from several other
characteristics \cite{r7,r8,r9,r10}. There are several nuances and
Compton scale phenomena like fuzzy spacetime
come in, but these do not concern us here, and can be found in the references.\\
We merely report that the linearized equations \cite{r2,r11} of
General Relativity, viz.,
\begin{equation}
g_{\mu \nu} = \eta_{\mu \nu} + h_{\mu \nu}, h_{\mu \nu} = \int
\frac{4T_{\mu \nu}(t-|\vec x - \vec x'|, \vec x')}{|\vec x - \vec
x'|} d^3 x'\label{e1}
\end{equation}
where as usual,
\begin{equation}
T^{\mu \nu} = \rho u^u u^v\label{e2}
\end{equation}
lead to the mass, spin, gravitational potential and charge of an
electron, if we work at the Compton scale (Cf.ref.\cite{r6} for
details). Let us now apply the macro Gravitoelectic and
Gravitomagnetic equations to the above case. Infact these
equations are (Cf.ref.\cite{r1}).
\begin{equation}
\nabla \cdot \vec E_g \approx -4\pi \rho, \nabla \times \vec E_g
\approx - \partial \vec H_g/\partial t, etc.\label{e3}
\end{equation}
\begin{equation}
\vec E_g = - \nabla \phi - \partial \vec A/\partial t, \quad \vec
H_g = \nabla \times \vec A\label{e4}
\end{equation}
\begin{equation}
\phi \approx - \frac{1}{2} (g_{00} + 1), \vec A_\imath \approx
g_{0 \imath},\label{e5}
\end{equation}
The subscripts $g$ in the equations (\ref{e3}), (\ref{e4}),
(\ref{e5}) are to indicate that the fields $E$ and $H$ in the
macro case do not really represent the Electromagnetic field, but
rather resemble them. Let us apply equation (\ref{e4}) to equation
(\ref{e1}), keeping in mind equation (\ref{e5}). We then get,
considering only the order of magnitude, which is what interests
us here, after some manipulation
\begin{equation}
|\vec H | \approx \int \frac{\rho V}{r^2} \bar r \approx \frac{m
V}{r^2}\label{e6}
\end{equation}
and
\begin{equation}
| \vec E | = \frac{mV^2}{r^2}\label{e7}
\end{equation}
$V$ being the speed.\\
In (\ref{e6}) and (\ref{e7}) the distance $r$ is much greater than
a typical Compton wavelength, to make the approximations
considered in
deriving the Gravitomagnetic and Gravitoelectric equations meaningful.\\
Remembering that we have
$$mVr \approx h,$$
the electric and magnetic fields in (\ref{e6}) and (\ref{e7}) now
become
\begin{equation}
|\vec H | \sim \frac{h}{r^3} , |\vec E | \sim
\frac{hV}{r^3}\label{e8}
\end{equation}
We now observe that (\ref{e8}) does not really contain the mass of
the
elementary particle. Could we get a further insight into this new force?\\
Indeed in the above linearized General Relativistic
characterisation of the electron, it turns out that the electron
can be represented by the Kerr-Newman metric (Cf.\cite{r6} for
details). This incidentally also gives the anomalous gyromatgnetic
ratio $g=2$. This result has recently been reconfirmed by Nottale
\cite{r12} from a totally different point of view, using scaled
relativity. It is well known that the Kerr-Newman field has extra
electric and magnetic terms (Cf.\cite{r13}),
both of the order $\frac{1}{r^3}$, exactly as indicated in (\ref{e8}).\\
It may be asked if there is any candidate as yet for the above
short range force. There is already one such candidate - the
inexplicable $B_{(3)}$ \cite{r14} force mediated by massive
photons and of short range, first detected in 1992 at Cornell and
since confirmed by subsequent experiments. (It differs from the
usual $B_{(1)}$ and $B_{(2)}$ fields of special relativity,
mediated as
they are, by massless photons.)\\
A Final Comment: It is quite remarkable that equations like
(\ref{e3}), (\ref{e4}) and (\ref{e5}) which resemble the equations
of electromagnetism, have in the usual macro considerations no
connection whatsoever with electromagnetism except in appearance.
This would seem to be a rather miraculous coincidence. In fact the
above considerations of the linearized General Relativistic theory
of the electron as also the Kerr-Newman metric formulation,
demonstrate that the resemblence to electromagnetism is not an
accident, because in this latter formulation, both
electromagnetism and gravitation arise from the metric (Cf.also
refs.\cite{r15,r6,r7,r8}).

\end{document}